\documentclass[aps,prb,twocolumn,showpacs,amsmath,amssymb,superscriptaddress]{revtex4}
\usepackage{graphicx}
\usepackage{amssymb}
\usepackage{natbib}

\begin{document}

\title{In-plane spin excitation anisotropy in the paramagnetic state of NaFeAs}
\author{Yu Song}
\affiliation{Department of Physics and Astronomy, Rice University, Houston, Texas 77005, USA}
\affiliation{ Department of Physics and Astronomy,
The University of Tennessee, Knoxville, Tennessee 37996-1200, USA }
\author{Louis-Pierre Regnault}
\affiliation{SPSMS-MDN, UMR-E CEA/UJF-Grenoble 1, INAC, Grenoble, F-38054, France}
\author{Chenglin Zhang}
\affiliation{Department of Physics and Astronomy, Rice University, Houston, Texas 77005, USA}
\affiliation{ Department of Physics and Astronomy,
The University of Tennessee, Knoxville, Tennessee 37996-1200, USA }
\author{Guotai Tan}
\affiliation{ Department of Physics and Astronomy,
The University of Tennessee, Knoxville, Tennessee 37996-1200, USA }
\author{Scott V. Carr}
\affiliation{Department of Physics and Astronomy, Rice University, Houston, Texas 77005, USA}
\affiliation{ Department of Physics and Astronomy,
The University of Tennessee, Knoxville, Tennessee 37996-1200, USA }
\author{Songxue Chi}
\affiliation{Quantum Condensed Matter Division, Oak Ridge National Laboratory, Oak Ridge, Tennessee 37831, USA
}
\author{A. D. Christianson}
\affiliation{Quantum Condensed Matter Division, Oak Ridge National Laboratory, Oak Ridge, Tennessee 37831, USA
}
\author{Tao Xiang}
\affiliation{Beijing National Laboratory for
Condensed Matter Physics, Institute of Physics, Chinese Academy of
Sciences, Beijing 100190, China}
\author{Pengcheng Dai}
\email{pdai@rice.edu}
\affiliation{Department of Physics and Astronomy, Rice University, Houston, Texas 77005, USA}
\affiliation{ Department of Physics and Astronomy,
The University of Tennessee, Knoxville, Tennessee 37996-1200, USA }
\affiliation{Beijing National Laboratory for
Condensed Matter Physics, Institute of Physics, Chinese Academy of
Sciences, Beijing 100190, China}

\date{\today}

\begin{abstract}
We use unpolarized and polarized inelastic neutron scattering to study low-energy spin excitations in NaFeAs, which exhibits
a tetragonal-to-orthorhombic lattice distortion
 at $T_s\approx 58$ K followed by a collinear antiferromagnetic (AF) order below
$T_N\approx 45$ K.
In the AF ordered state ($T<T_N$), spin waves are entirely $c$-axis polarized below $\sim$10 meV,
exhibiting a gap of $\sim4$ meV at the AF
zone center and disperse to $\sim$7 meV near the $c$-axis
AF zone boundary.
On warming to the paramagnetic state with orthorhombic lattice
distortion ($T_N<T<T_s$), spin excitations become anisotropic within the FeAs plane.
Upon further warming to
the paramagnetic tetragonal state ($T>T_s$), spin excitations become more isotropic.
Since similar magnetic anisotropy is also observed in the paramagnetic tetragonal phase
of superconducting BaFe$_{1.904}$Ni$_{0.096}$As$_2$, 
our results suggest that the spin excitation anisotropy in superconducting iron pnictides originates from similar anisotropy already present in their
parent compounds.
\end{abstract}

\maketitle

\section{Introduction}
The parent compounds of iron pnictide superconductors are semi-metallic antiferromagnets exhibiting
a tetragonal to orthorhombic lattice distortion at temperature $T_s$ followed by a
paramagnetic to antiferromagnetic (AF) phase transition at
$T_N$ \cite{kamihara,cruz,qhunag,mgkim,cwchu,slli,ysong,dai}.  The magnetic structure is collinear with the ordered moment aligned antiferromagnetically
along the $a$-axis of the orthorhombic lattice [Fig. 1(a)].  From transport \cite{fisher}, resonant ultrasound \cite{fernandes10},
angle-resolved photoemission spectroscopy (ARPES) \cite{myi,yzhang,myi12}, inelastic neutron scattering \cite{harriger},
magnetic torque \cite{kasahara}, and scanning tunneling microscopy (STM) \cite{tmchuang,allan,pasupathy} measurements,
it is clear that iron pnictides have electronic anisotropy above $T_N$ implying an underlying electronic nematic phase \cite{fradkin}.
However, the microscopic origin of the observed electronic anisotropy is still an issue of debate.  In both the
strong and weak coupling limits, the electronic anisotropy could be caused by a spin nematic phase (spin anisotropy)
in the paramagnetic orthorhombic phase above $T_N$ but below $T_s$ \cite{jphu,fernandes1}.  Alternatively,
orbital ordering in the paramagnetic orthorhombic state may also induce the observed electronic anisotropy
 \cite{cclee,kruger,lv,ccchen,valenzeula}.  Although transport \cite{jhchu10} and X-ray diffraction \cite{ruff} experiments in magnetic fields
 on electron-doped BaFe$_{2-x}$Co$_{x}$As$_2$ reveal clear evidence for anisotropic in-plane
static spin susceptibility ($\chi_a\neq \chi_b$) in the paramagnetic orthorhombic state ($T_N\leq T \leq T_s$), it is still unclear if such
in-plane susceptibility anisotropy is field-induced or intrinsic to these materials at zero field.  Furthermore, recent STM \cite{allan}
and transport measurements \cite{nakajima,ishida} suggest that the resistivity anisotropy in Co-doped BaFe$_2$As$_2$ arises from Co-impurity scattering,
and is not an intrinsic property of these materials.

\begin{figure}[t]
\includegraphics[scale=.35]{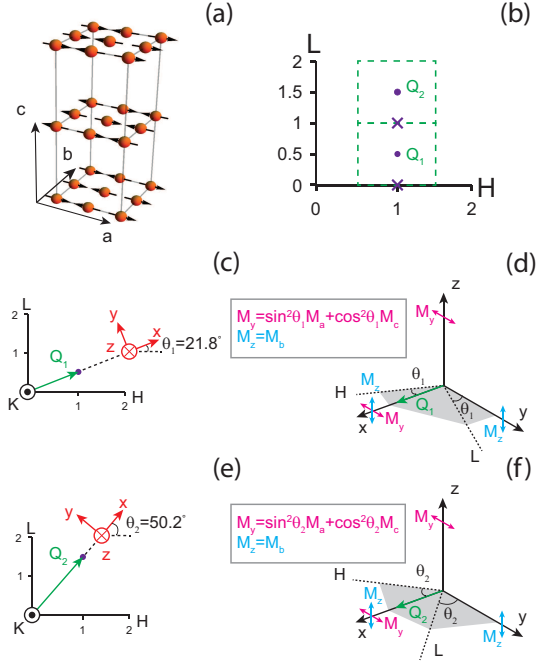}
\caption{
(Color online) (a) The orthorhombic chemical unit cell of NaFeAs (enclosed by grey lines, two orthorhombic chemical unit cells are stacked along the $c$-axis), the black arrows indicate directions of the ordered moments and orange spheres represent Fe atoms, Na and As atoms are not shown. (b) Points in reciprocal space probed in this work.
The points represent magnetic zones centers ($ L=0.5,1.5$) while the crosses represent magnetic zone boundaries ($L=0,1$) along the $(1,0,L)$ direction.
The dashed green lines enclose the magnetic Brillouin zone.  The relationship between the neutron polarization directions ($x,y,z$) and the scattering plane are shown in (c) for ${\bf Q}= (1,0,0.5)$ and in (e) for $(1,0,1.5)$.  The angle between $x$-direction and the $H$-axis is denoted as $\theta_{1}$ for $(1,0,0.5)$ and $\theta_{2}$ for $(1,0,1.5)$.  The magnetic components measured in each channel and their relationships to the magnetic components projected onto the crystal axes are shown in (d) for ${\bf Q}= (1,0,0.5)$ and in (f) for $(1,0,1.5)$.  Only SF channels are measured in this work.  $\sigma_{x}^{\rm SF}$ contains both $M_y$ and $M_z$ magnetic components, whereas only $M_y$ and $M_z$ contribute to $\sigma_{z}^{\rm SF}$ and $\sigma_{y}^{\rm SF}$, respectively. The relationships between the magnetic components along the neutron
polarization directions ($M_x$, $M_y$ and $M_z$) and crystal axes ($M_a$, $M_b$, and $M_c$) are shown in the boxes in (d) and (f) for ${\it L}=0.5$ and ${\it L}=1.5$ respectively.
}
\end{figure}

In this article, we use unpolarized and polarized neutron scattering to study spin waves and paramagnetic spin excitations in NaFeAs \cite{slli}. In contrast to BaFe$_2$As$_2$,
where the orthorhombic lattice distortion ($T_s$) and AF order ($T_N$) occur at similar temperatures \cite{qhunag,mgkim}, NaFeAs has clearly separated structural and magnetic
phase transitions at $T_s\approx 58$ K and $T_N\approx 45$ K, respectively \cite{slli,ysong}.
Below the AF ordering temperature, the iron spins in NaFeAs order antiferromagnetically 
along the $a$-axis of the orthorhombic structure and ferromagnetically along the $b$-axis \cite{slli,ysong}.
In the low-temperature AF ordered state, NaFeAs forms randomly distributed 
orthorhombic twin domains rotated 90$^\circ$ apart similar to BaFe$_2$As$_2$ \cite{harriger}. 
Low energy spin excitations in iron pnictides are centered around
the AF ordering wave vectors ${\bf Q_{AF}}=(\pm1,0)$ and $(0,\pm1)$ corresponding to the two sets of domains, thus 
allowing spin excitations polarized along different crystallographic axes to be determined in a twinned sample\cite{olly,qureshi,mengshu,clzhang,hluo}.
In the AF ordered state ($T<T_N$) of NaFeAs, our unpolarized neutron scattering measurements 
find that spin waves are gapped below $\sim4$ meV at the AF
zone center and disperse to $\sim$7 meV near the $c$-axis AF zone boundary.  
Similar to BaFe$_2$As$_2$ \cite{qureshi},
neutron polarization analysis indicates that spin waves in NaFeAs are entirely $c$-axis polarized for energies below $\sim$10 meV.
On warming the system to the paramagnetic orthorhombic state above $T_N$, the $c$-axis polarized spin waves become
anisotropic paramagnetic scattering. 
By carefully measuring wave vector dependence of the paramagnetic scattering, we show that
the magnetic response $M$ ($M$ is related to the imaginary part of the dynamic susceptibility, $\chi^{\prime\prime}({\bf Q},E)$, via the Bose factor  $M=\chi^{\prime\prime}({\bf Q},E)/[1-\exp(-{E/{k_{B}T}})]$.) at the AF wave vector 
and $E=6$ meV has in-plane anisotropy with
$M_a \geq M_b$ in the
 paramagnetic orthorhombic state ($T_N\leq T\leq T_s$) of NaFeAs.
Such anisotropy becomes much weaker above $T_s$ and spin excitations become nearly isotropic.
Since the in-plane anisotropic paramagnetic spin excitations in NaFeAs are similar to those observed in 
 the tetragonal phase of superconducting BaFe$_{1.906}$Ni$_{0.096}$As$_2$ \cite{hluo}, 
the spin excitation anisotropy in superconducting iron pnictides originates from similar anisotropy already present in the
parent compound.  While these results suggest 
the presence of a spin nematic state in the paramagnetic orthorhombic phase of NaFeAs, they may also be consistent with 
anisotropic critical fluctuations due to a single-ion anisotropy in the orthorhombic phase \cite{collin}.

\begin{figure}[t]
\includegraphics[scale=.5]{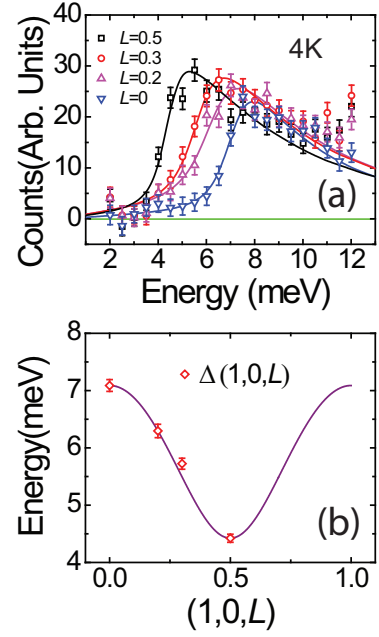}
\caption{
(Color online) (a) Constant-${\bf Q}$ scans at $(1,0,{\it L})$ at 4 K for ${\it L}=0, 0.2, 0.3$ and 0.5 with unpolarized neutrons after background has been subtracted. Solid lines are fits to the response of a damped harmonic oscillator convolved with instrumental resolution with low-energy spin waves modeled as $E({\bf Q})=\sqrt{\Delta(1,0,{\it L})^2+v_a^2h^2+v_b^2k^2}$, where $(h,k)=(H,K)-(1,0)$ \cite{matan}. Data points with $E<11$meV are used in the fits.
(b) Dispersion of the low energy spin waves along the $(1,0,{\it L})$ direction, points are fitted gap values $\Delta(1,0,{\it L})$, the purple solid line is the dispersion from linear spin wave theory using effective exchange couplings from Ref. \cite{clzhang13}.
 }
 \end{figure}

\section{Experimental Results}

Figure 1(a) shows the collinear AF structure of NaFeAs with orthorhombic lattice parameters $a=5.589$, $b= 5.569$ and $c=6.991$ \AA\ \cite{slli}.
We define momentum transfer ${\bf Q}$ in three-dimensional reciprocal space in \AA$^{-1}$ as $\textbf{Q}=H\textbf{a}^\ast+K\textbf{b}^\ast+L\textbf{c}^\ast$,
where
$H$, $K$, and $L$ are Miller indices and
${\bf a}^\ast=\hat{{\bf a}}2\pi/a$, ${\bf b}^\ast=\hat{{\bf b}}2\pi/b$, ${\bf c}^\ast=\hat{{\bf c}}2\pi/c$.
About 5 grams of single crystals of NaFeAs were coaligned in the $[H,0,L]$ scattering plane [Fig. 1(b)].  In this notation,
the AF Bragg peaks and zone centers occur at $[1,0,L]$ with $L=0.5,1.5,\cdots$, while the AF zone boundaries along the $c$-axis occur at $L=0,1,2,\cdots$ \cite{slli}.
Our polarized neutron scattering experiments were
carried out using the IN22 thermal triple-axis spectrometer at the Institut
Laue-Langevin, Grenoble, France \cite{olly}.  Consistent with previous notation \cite{olly}, we define the neutron polarization directions
along ${\bf Q}$ as $x$, perpendicular to ${\bf Q}$ but in the scattering plane
as $y$, and perpendicular to ${\bf Q}$ and the scattering plane as $z$, respectively [Figs. 1(c) and 1(e)].
Since neutron scattering is only sensitive to magnetic scattering components perpendicular to the momentum transfer
\textbf{Q}, one can probe magnetic responses within the $y-z$ plane ($M_y$ and $M_z$) [Fig. 1(c)].

Within the scattering plane,
the measured magnetic response at ${\bf Q}$ give $M_y=\sin^2\theta M_a+\cos^2\theta M_c$ and $M_z=M_b$,
where the angle between ${\bf Q}$ and $[H,0,0]$ is $\theta$, and $M_a$, $M_b$, and $M_c$ are spin excitation intensities along the
orthorhombic $a$-, $b$-, and $c$-axis directions, respectively \cite{olly}.  This is related to
the observed neutron spin-flip (SF) scattering cross sections for different neutron polarization directions
$\sigma_x^{\rm SF}$, $\sigma_y^{\rm SF}$, and $\sigma_z^{\rm SF}$ via:
\begin{equation}
\left\{
\begin{array}{cc}
\sigma_x^{\rm SF} =\frac{R}{R+1}(\sin^2\theta M_a+\cos^2\theta M_c)+\frac{R}{R+1}M_b+B,\\
\\[1pt]
\sigma_y^{\rm SF} =\frac{1}{R+1}(\sin^2\theta M_a+\cos^2\theta M_c)+\frac{R}{R+1}M_b+B,\\
\\[1pt]
\sigma_z^{\rm SF} =\frac{R}{R+1}(\sin^2\theta M_a+\cos^2\theta M_c)+\frac{1}{R+1}M_b+B\\
\end{array}
\right.
\end{equation}
where $R$ is the flipping ratio for
non-spin-flip (NSF) and SF scattering ($R=\sigma_{\rm Bragg}^{\rm NSF}/\sigma_{\rm Bragg}^{\rm SF}\approx 15$) and $B$ is the background scattering that may be larger than the magnetic scattering.
Therefore, to conclusively determine the magnetic anisotropy along the three
crystallographic
directions $M_a$, $M_b$, and $M_c$
of the orthorhombic lattice, we need to
measure neutron SF scattering at two or more equivalent AF wave vectors with different
angle $\theta$ between the wave vector ${\bf Q}$ and $[H,0,0]$ \cite{hluo}.  To compare the estimated $M_a$, $M_b$, and $M_c$ at different wave vectors, we need to consider in addition
the differences in the magnetic form factor $F({\bf Q})$ and instrumental resolution $r$.
For example, at the AF zone center [1,0,L] ($L=0.5,1.5,\cdots$) positions, by combining the cross sections measured at
${\bf Q}=(1,0,0.5)$ and ${\bf Q}=(1,0,1.5)$ for a particular energy transfer ($\sigma_x^{\rm SF}$, $\sigma_y^{\rm SF}$ and $\sigma_z^{SF}$ at both wave vectors), we have five quantities
($M_a$, $M_b$, $M_c$ and $B$ at the two wave vectors)
to be determined from up to six cross sections [Figs. 1(c)-1(f)].  We either measured all six cross sections and used the over-determination to improve estimates of $M_a$, $M_b$ and $M_c$ [Fig. 3(a)-3(d)] or measured five cross sections which uniquely determines the magnetic response [Fig. 4(c)-4(d)].
A similar procedure is used to analyze data for the zone boundary along the $c$-axis ($L=0,1$).

In previous unpolarized neutron scattering measurements of spin waves in Na$_{0.9}$FeAs,
the onset of spin gap at the AF zone center ${\bf Q}=(1,0,1.5)$ is observed at approximately $\sim$10 meV \cite{jtpark12}. Figure 2 summarizes our unpolarized neutron measurements on NaFeAs using
HB-3 triple-axis spectrometer at the High-Flux-Isotope Reactor (HFIR), Oak Ridge National Laboratory.
The monochromator, analyzer, and filters are all pyrolytic graphite.
The collimations are $48^{\prime}$-$40^{\prime}$-sample-$40^{\prime}$-$120^{\prime}$ with final neutron
energy fixed at $E_f=14.68$ meV.
Our energy scan
at the AF zone center wave vectors ${\bf Q}=(1,0,0.0.5)$ reveals a clear spin gap of $\sim$4 meV [Fig. 2(a)],
much smaller than that of Na$_{0.9}$FeAs \cite{jtpark12}.   
Upon moving the wave vectors to
${\bf Q}=(1,0,0.3)$,  $(1,0,0.2)$, and $(1,0,0)$,
the spin gap changes to $\sim$7 meV at the $c$-axis AF zone boundary position with $L=0$ [Fig. 2(a)].
To understand these data, we fit the spin wave spectra with a damped harmonic oscillator,
$\chi^{\prime\prime}({\bf Q},E)=A E_0({\bf Q})\Gamma E/[(\Gamma E)^2+(E_0^2({\bf Q})-E^2)^2]$, 
convolved with instrumental resolution similar to 
previous work \cite{matan}.  The spin wave dispersion is 
$E({\bf Q})=\sqrt{\Delta(1,0,{\it L})^2+v_a^2h^2+v_b^2k^2}$, where $\Delta(1,0,{\it L})$ is the spin gap
value, $v_a$ and $v_b$ are spin wave velocities along the $a$- and $b$-axes, respectively.  
The solid lines in Fig. 2(a) show the fits 
using spin wave velocities of NaFeAs obtained from high-energy time-of-flight measurements \cite{clzhang13}.
Figure 2(b) shows the $c$-axis dispersion of the spin waves.
Given the almost identical $T_N$ and $T_s$ between our NaFeAs samples \cite{ysong} and Na$_{0.9}$FeAs \cite{jtpark12},  
it is unclear why their spin gap values are so different.

\begin{figure}[t]
\includegraphics[scale=.35]{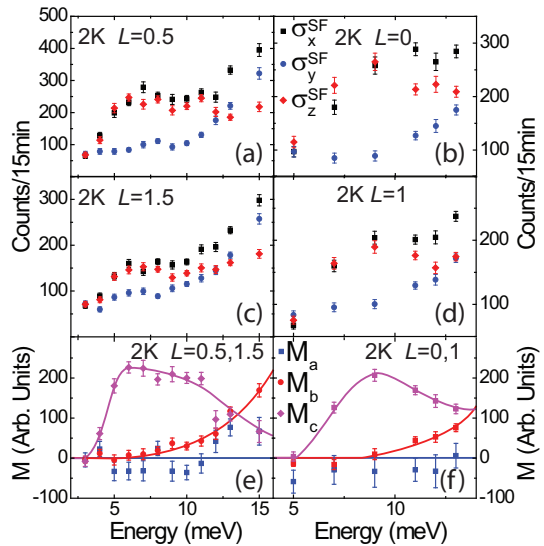}
\caption{
Color online) Constant-${\bf Q}$ scans in the AF ordered state ($T=2$ K) at ${\bf Q}=(1,0,L)$ with ${\it L}=0, 0.5, 1$ and 1.5 are shown in (b), (a), (d), and (c), respectively. All three spin-flip channels are measured, denoted as $\sigma_{x}^{\rm SF}$, $\sigma_{y}^{\rm SF}$ and $\sigma_{z}^{\rm SF}$ for the three different neutron polarization directions.  From the measured cross sections at $(1,0,0.5)$ and $(1,0,1.5)$, the magnetic components at the magnetic zone center along the crystallographic axes $M_a$, $M_b$, and $M_c$ are determined and plotted in (e).  Similarly, (f) shows the magnetic components along different crystallographic axes for magnetic zone boundaries at ${\bf Q}=(1,0,0)$ and $(1,0,1)$.
The solid lines in (e) and (f) are guides to the eye.
 }
\end{figure}

\begin{figure}[t]
\includegraphics[scale=.32]{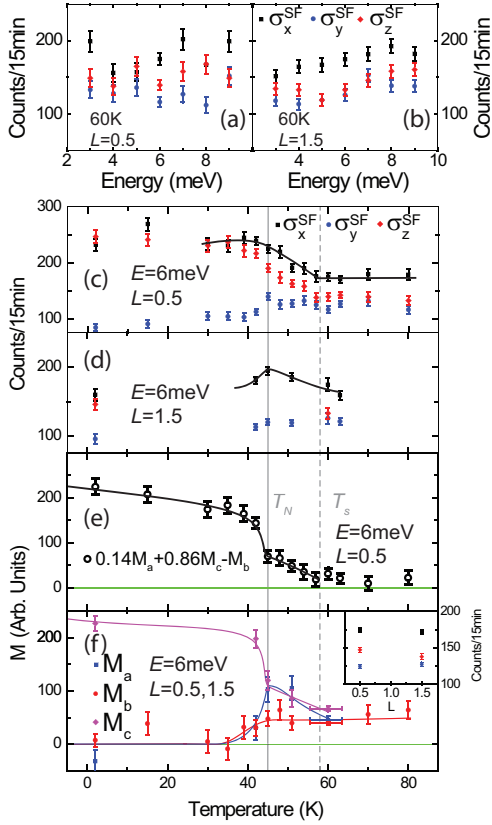}
\caption{
(Color online) Constant-${\bf Q}$ scans at $T=60$ K for $(1,0,0.5)$ and $(1,0,1.5)$ are shown in (a) and (b), respectively.  Temperature dependence of the
three spin-flip cross sections for the AF zone centers $(1,0,0.5)$ and $(1,0,1.5)$ are shown in (c) and (d). (e) The difference of the magnetic components $M_y$ and $M_z$ for ${\it L}=0.5$ determined from $\sigma_{z}^{\rm SF}$ and $\sigma_{y}^{\rm SF}$ in (c).  The temperature dependence of $M_a$, $M_b$ and $M_c$ at the magnetic zone center are determined and plotted in (f). In cases where only cross sections for ${\it L}=0.5$ are measured, only $M_b$ can be determined, when ${\sigma_{x}^{\rm SF}}$ and ${\sigma_{y}^{\rm SF}}$ for ${\it L}=1.5$ are in addition measured, all three components along crystal axes can be determined, when all three channels are measured for both ${\it L}=0.5$ and ${\it L}=1.5$, the over-determination is used to improve estimates of $M_a$, $M_b$ and $M_c$.  The point at $T\approx60$ K is obtained by combining raw data from temperatures in the range indicated by horizontal bars and the constant-${\bf Q}$ scans in (a) and (b) from $E=4$-8 meV, the combined $\sigma_{x}^{\rm SF}$, $\sigma_{y}^{\rm SF}$ and $\sigma_{z}^{\rm SF}$ for ${\it L}=0.5$ and ${\it L}=1.5$ are shown in the inset.  The solid vertical grey line through (c), (d), (e) and (f) marks the magnetic transition temperature $T_{N}$ whereas the dashed grey line marks the structural transition temperature $T_{s}$. The solid lines in (e) and (f) are guides to the eye.
 }
\end{figure}

Figure 3 shows the energy scans at the AF zone center ${\bf Q}=(1,0,0.5), (1,0,1.5)$ and zone boundary $(1,0,0), (1,0,1)$
with different neutron polarizations in the low-temperature AF ordered state.
At the AF wave vectors ${\bf Q}=(1,0,0.5)$ [Fig. 3(a)],
we find a clear spin gap of $\sim$4 meV consistent with unpolarized
data of Fig. 2.  Furthermore,
$\sigma_x^{\rm SF}$ are similar to $\sigma_z^{\rm SF}$ for $E\leq 11$ meV
while $\sigma_x^{\rm SF} > \sigma_y^{\rm SF} > \sigma_z^{\rm SF}$ for $E> 11$ meV.
Similar results are obtained at ${\bf Q}=(1,0,1.5)$ [Fig. 3(c)].  In order to obtain $M_a$, $M_b$, and $M_c$, we assume Fe$^{2+}$ magnetic form factor and correct for the instrumental resolution factor $r$ at these two wave vectors.
The obtained $M_a$, $M_b$ and $M_c$ are shown in Fig. 3(c).
We see that spin waves in NaFeAs are transverse to the ordered moment direction and
almost entirely $c$-axis polarized with $M_a\approx M_b \approx 0$
for $E\leq 10$ meV [Fig. 1(e)].  This is similar to the low temperature spin waves in BaFe$_2$As$_2$ \cite{qureshi}.
For spin wave energies above 10 meV, we see a dramatic reduction in $M_c$ and a corresponding increase in $M_b$.
Most surprisingly, there appears to be a small, but nonzero $M_a$ around 12 meV, suggesting the possible presence of longitudinal
spin excitations.  This is disallowed for spin waves in a classical local moment Heisenberg Hamiltonian, but may be
present in NaFeAs due to itinerant electrons \cite{dai}. 
Figures 3(b), 3(d) show similar results at the AF zone boundary  ${\bf Q}=(1,0,0)$ and $(1,0,1)$.  The energy dependence of
$M_a$, $M_b$, and $M_c$ are shown in Fig. 3(f).  Again, spin waves are entirely polarized along the $c$-axis for energies
below 10 meV and have no longitudinal component for the probed energy range. We note that recent polarized neutron scattering experiments on 
BaFe$_2$As$_2$ have conclusively established the presence of longitudinal spin-wave excitations in the AF ordered phase \cite{cwang}.

Figures 4(a) and 4(b) plot the energy scans in the paramagnetic tetragonal state ($T=60$ K) at
the AF wave vectors ${\bf Q}=(1,0,0.5)$ and $(1,0,1.5)$, respectively.
From Eq. (1), we note that isotropic paramagnetic scattering should imply 
$\sigma^{\rm SF}_x/2 \approx \sigma^{\rm SF}_y \approx \sigma^{\rm SF}_z$ in the limit of large $R$ and 
negligible background scattering ($B\rightarrow 0$).
This is clearly not the case in Figs. 4(a) and 4(b), thus suggesting the presence of finite background scattering.
In this case, one can estimate the magnetic anisotropy using $\sigma_z^{\rm SF}-\sigma_y^{\rm SF}=
(\sin^2\theta M_a+\cos^2\theta M_c-M_b)(R-1)/(R+1)
\propto M_y-M_z$ to eliminate the influence of background. Since data in 
Figs. 4(a) and 4(b) suggest $\sigma_x^{\rm SF}>\sigma_y^{\rm SF}\approx \sigma_z^{\rm SF}$, 
there should be weak magnetic anisotropy between $M_y$ and $M_z$ in 
the paramagnetic tetragonal state.
To probe the evolution of spin excitations in the paramagnetic orthorhombic ($T_N\leq T\leq T_s$) and tetragonal ($T>T_s$) phases,
we study the temperature dependence of spin excitations at $E=6$ meV from 2 K to 80 K.
Figures 4(c) and 4(d) show the raw data obtained for
different neutron polarizations at the AF wave vectors
${\bf Q}=(1,0,0.5)$ and $(1,0,1.5)$, respectively, across the AF ordered, paramagnetic orthorhombic and tetragonal phases.
In the low-temperature AF ordered state ($T=2$ K), we find $\sigma_x^{\rm SF}\approx \sigma_z^{\rm SF}$
consistent with data in Fig. 3(a).  On warming to the
paramagnetic orthorhombic state [$T_N\leq T\leq T_s$, temperatures between the solid and dashed vertical lines
in Figs. 4(c)-4(f)], we have $\sigma_x^{\rm SF} > \sigma_z^{\rm SF} > \sigma_y^{\rm SF}$.  This means
anisotropic  
spin excitations at ${\bf Q}=(1,0,0.5)$ with $M_y-M_z=0.14M_a+0.86M_c-M_b>0$ [Fig. 1(c)].
Finally, on warming to temperatures above $T_s$ [Figs. 4(c) and 4(d)],
spin excitations 
satisfy $\sigma_x^{\rm SF}> \sigma_z^{\rm SF} \approx \sigma_y^{\rm SF}$ 
and are consistent with data in Figs. 4(a) and 4(b).  Therefore, spin excitations in the
paramagnetic tetragonal state of NaFeAs 
are more isotropic with 
$M_y-M_z=0.14M_a+0.86M_c-M_b\approx 0$.

Given the experimental evidence for anisotropic spin excitations in the paramagnetic orthorhombic phase of
NaFeAs, it is important to determine its anisotropy along the crystallographic axes.
In the AF ordered state, the low-energy spin waves from the two 90$^\circ$ rotated twin domains 
are centered around wave vectors ${\bf Q_{AF}}=(\pm1,0)$ and $(0,\pm1)$, respectively, in reciprocal space.  
Therefore, low-energy spin waves from the $(\pm1,0)$ domain are not mixed with those from 
the $(0,\pm1)$ domain.  However, in the paramagnetic orthorhombic state, spin excitations at the 
wave vector $(\pm1,0)$ may be mixed with paramagnetic excitations from the domain associated with $(0,\pm1)$, thus complicating 
the neutron polarization analysis.
The key question is whether there are strong paramagnetic scattering at the wave vector $(0,\pm1)$
in a completely detwinned sample associated with the AF 
wave vectors $(\pm1,0)$.  Although such measurement for NaFeAs is unavailable, we note that 
neutron scattering experiments on a nearly 100\% mechanically detwinned BaFe$_2$As$_2$ reveal that
spin excitations in the paramagnetic tetragonal state are still centered mostly at ${\bf Q_{AF}}=(\pm1,0)$ $\sim$20 K above the 
AF and structural transition temperatures \cite{xylu13}.  Therefore, spin excitations of NaFeAs at the wave vectors  
${\bf Q_{AF}}=(\pm1,0)$ may also have little contribution from those
associated with the $(0,\pm1)$ domain 
in the paramagnetic orthorhombic state.  Assuming this is indeed the case, we can carry out
neutron polarization analysis in the paramagnetic orthorhombic phase similar to the AF ordered state.  
Figure 4(e) shows the temperature dependence of $0.14M_a+0.86M_c-M_b$ for ${\bf Q}=(1,0,0.5)$ and $E=6$ meV.
The data show a clear kink at $T_N$ and positive scattering below $T_s$, meaning $M_b<0.14M_a+0.86M_c$.   The temperature dependence of
$M_a$, $M_b$, and $M_c$ shown in Fig. 4(f) are determined from
combining the data in Figs. 4(a)-4(d).   Inspection of the Figure reveals clear 
in-plane magnetic anisotropy with $M_a>M_b$ in the paramagnetic orthorhombic phase that
becomes much smaller in the paramagnetic tetragonal phase [Fig. 4(f)].

Our results suggest that
whereas $M_b$, $M_c$ evolve smoothly across $T_N$, $M_a$ peaks around $T_N$ indicating 
divergent longitudinally polarized critical magnetic fluctuations. 
Both $\sigma_x^{\rm SF}$ and unpolarized neutron scattering measures $M_y+M_z$, which is $0.14M_a+0.86M_c+M_b$ for ${\bf Q}=(1,0,0.5)$ and $0.59M_a+0.41M_c+M_b$ for ${\bf Q}=(1,0,1.5)$.  If $M_a$ dominates the critical fluctuations near $T_N$, one would expect a stronger peak due to critical fluctuations at $(1,0,1.5)$ than at $(1,0,0.5)$ since 
magnetic structural factor is larger at $(1,0,1.5)$ \cite{slli}.
Comparison of our data measured at these two wave vectors [Fig. 4(c)-(d)] and unpolarized neutron scattering results in Ref. \cite{jtpark12} suggest this is indeed the case.
For a classical Heisenberg magnetic system with an Ising anisotropy term, one would expect diverging longitudinally polarized spin excitations at $T_N$ 
consistent with our observation \cite{collin}.

In previous polarized neutron measurements on parent compound BaFe$_2$As$_2$ \cite{qureshi}, it was found that the in-plane polarized spin waves exhibit a larger gap than the $c$-axis polarized ones.  However,
the spin anisotropy immediately disappears in the paramagnetic tetragonal state above
$T_N$ and $T_s$ \cite{qureshi}.  In addition, while 
$\sigma_x^{\rm SF}-\sigma_z^{\rm SF} \propto M_b$ diverges at $T_N$, 
$\sigma_x^{\rm SF}-\sigma_y^{\rm SF} \propto(\sin^2\theta M_a+\cos^2\theta M_c)$ peaks at a temperature slightly below $T_N$ \cite{qureshi}.
This is different from the results in Fig. 4.
Since $T_N$ and $T_s$ occur at almost the same temperature in BaFe$_2$As$_2$ \cite{qhunag,mgkim}, it is unclear whether
the system also has magnetic anisotropy in the paramagnetic orthorhombic phase.  
The discovery of an in-plane spin excitation anisotropy in
the paramagnetic orthorhombic phase of NaFeAs suggests the presence of a strong spin-orbit coupling in such a state \cite{olly,boothroyd,prokes,steffens,clzhang}.
However, we cannot distinguish if such anisotropy is a sole manifestation of spin nematicity or a consequence of the orbital ordering \cite{cclee,kruger,lv,ccchen,valenzeula}.  
In a recent X-ray diffraction experiment under pulsed magnetic fields, in-plane field-induced static
susceptibility anisotropy with $\chi_b> \chi_a$ in the AF ordered state is found to extend to the paramagnetic orthorhombic phase of
electron-underdoped Ba(Fe$_{1-x}$Co$_x$)$_2$As$_2$ \cite{ruff}.  This is 90$^\circ$ rotated from the 
in-plane dynamic susceptibility anisotropy ($M_a>M_b$ or $\chi_a^{\prime\prime}>\chi_b^{\prime\prime}$) in 
the paramagnetic orthorhombic state of NaFeAs.  While
the static susceptibility anisotropy remains unchanged from the AF ordered phase to the paramagnetic orthorhombic phase in 
Ba(Fe$_{1-x}$Co$_x$)$_2$As$_2$ \cite{ruff},
there is a dramatic switch over of the spin excitation anisotropy
across $T_N$ in NaFeAs, changing from the entirely $c$-axis polarized spin waves ($M_c\gg M_a\approx M_b\approx 0$)
in the AF ordered phase to
$M_a\geq M_c>M_b$ in the paramagnetic orthorhombic state.  Finally, the spin anisotropy becomes much smaller in the
paramagnetic tetragonal phase.  We note that the static in-plane susceptibility anisotropy observed in transport \cite{jhchu10} and X-ray diffraction experiments \cite{ruff} 
occurs at the zero wave vector, while the dynamic susceptibility anisotropy in NaFeAs 
is at the AF wave vectors ${\bf Q_{AF}}=(\pm1,0)$.

\section{Discussion and Conclusions}

To qualitatively understand these results, we note that
the orthorhombic lattice distortion below $T_s$ lifts the
degeneracy between Fe $d_{xz}$ and $d_{yz}$ orbitals and leads to a ferro-orbital order with more doubly-occupied Fe $d_{xz}$ orbitals and more singly-occupied Fe $d_{yz}$ orbitals \cite{cclee}.  In the case of NaFeAs, the Fermi surfaces evolve dramatically from the paramagnetic tetragonal state to the AF ordered state and the splitting of the $d_{xz}$ and $d_{yz}$ orbitals starts to 
occurs at a temperature above $T_s$
\cite{yzhang}. 
The angular momentum of $d_{yz}$ orbitals lies along the $a$-axis direction, which pins the ordering spin moment along the $a$-axis
via spin-orbit interaction.  This is indeed observed experimentally as shown in Fig. 1(a) \cite{qhunag}.
Although the electronic structure undergoes an orbital-dependent reconstruction in the nematic state above $T_s$, 
primarily involving the splitting of $d_{xz}$- and $d_{yz}$-dominated bands, 
the splitting mostly occurs in the temperature range 
above $T_N$, and there are only small changes across $T_N$ \cite{yzhang}.  
This is different from the temperature dependent spin dynamic susceptibility across $T_N$ [Fig. 4(f)], but consistent with 
the notion that ferro-orbital ordering involving $d_{xz}$- and $d_{yz}$ bands 
plays a minor role \cite{yzhang}.  Therefore, it is more likely the observed dynamic susceptibility anisotropy is a manifestation of dynamic spin nematicity coupled with orbital ordering.
For a typical second order AF phase transition, critical spin fluctuations should
exhibit a peak at $T_N$ \cite{leland09}.
From Fig. 4(f), we see that spin excitations in NaFeAs are dominated by the longitudinal fluctuations ($M_a$ or
$\chi^{\prime\prime}_a$) near $T_N$, consistent with diverging longitudinally polarized spin excitations in a Heisenberg antiferromagnet with Ising spin anisotropy \cite{collin}. 
Given the small lattice distortion in the paramagnetic orthorhombic phase of NaFeAs \cite{ysong}, it is unlikely such an anisotropy term could arise from the lattice distortion. 
This is consistent with the fact that similar in-plane spin excitation anisotropy has also been observed in the paramagnetic tetragonal phase of superconducting 
BaFe$_{1.904}$Ni$_{0.096}$As$_{2}$ \cite{hluo}.  Instead, the observed in-plane spin excitation anisotropy is more likely to arise from 
 orbital ordering or anisotropic exchange interactions due to spin nematicity. 
 
In summary, we have discovered that low-energy spin waves in NaFeAs are entirely $c$-axis polarized for energies below $\sim$10 meV.
On warming the system across $T_N$ to the paramagnetic orthorhombic state, a clear in-plane anisotropy develops in the low-energy spin excitations
spectra, resulting $M_a\geq M_c>M_b$.  Finally, spin excitations become essentially isotropic in the paramagnetic tetragonal phase above $T_s$.
These results indicate that the spin excitation anisotropy in superconducting iron pnictides originates from similar anisotropy already present in their
parent compounds, and suggest the presence of a spin nematic phase in the paramagnetic
orthorhombic state of NaFeAs.

\section{ACKNOWLEDGMENTS}

We thank Haifeng Li and T. Netherton for experimental assistance, Jiangping Hu and Fa Wang for helpful discussions.
The single crystal growth efforts and neutron scattering work at UTK and Rice are supported by the US DOE, BES,
through contract DE-FG02-05ER46202. Work at IOP is supported by MOST (973 Project: 2012CB82400).
The work at
the HFIR, ORNL, was sponsored by the
Scientific User Facilities Division, BES, U.S. DOE.


\begin{thebibliography}{}

\bibitem{kamihara} Y. Kamihara, T. Watanabe, M. Hirano, and H. Hosono, J. Am. Chem. Soc. \textbf{130}, 3296-3297 (2008).

\bibitem{cruz} C. de la Cruz, Q. Huang, J. W. Lynn, J. Y. Li, W. Ratcliff II, J. L. Zarestky, H. A. Mook, G. F. Chen, J. L. Luo, N. L. Wang, and P. C. Dai, Nature (London) \textbf{453},899 (2008).

\bibitem{qhunag} Q. Huang, Y. Qiu, W. Bao, M. A. Green, J. W. Lynn, Y. C.
Gasparovic, T. Wu, G. Wu, and X. H. Chen, Phys. Rev. Lett. {\bf 101},
257003 (2008).

\bibitem{mgkim} M. G. Kim, R. M. Fernandes, A. Kreyssig, J. W. Kim, A. Thaler, S. L. Bud'ko, P. C. Canfield, R. J. McQueeney, J. Schmalian, and A. I. Goldman, Phys. Rev. B {\bf 83}, 134522 (2011).

\bibitem{cwchu} C. W. Chu, F. Chen, M. Gooch, A.M. Guloy, B. Lorenz, B. Lv, K. Sasmal, Z. J. Tang, J. H. Tapp, Y. Y. Xue, Physica C \textbf{469}, 326
(2009).

\bibitem{slli} S. L. Li, C. de la Cruz, Q. Huang, G. F. Chen, T.-L. Xia, J. L. Luo, N. L. Wang, and P. C. Dai, Phys. Rev. B \textbf{80}, 020504(R)
(2009).

\bibitem{ysong} Yu Song, Scott V. Carr, Xingye Lu, Chenglin Zhang, Zachary C. Sims, N. F. Luttrell, Songxue Chi, Yang Zhao, Jeffrey W. Lynn, Pengcheng Dai, Phys. Rev. B {\bf 87}, 184511 (2013).

\bibitem{dai} P. C. Dai, J. P. Hu, and E. Dagotto, Nature Phys. {\bf 8}, 709 (2012).

\bibitem{fisher} I. R. Fisher, L. Degiorgi, and Z. X. Shen, Rep. Prog. Phys. {\bf 74}, 124506 (2011).

\bibitem{fernandes10} R. M. Fernandes, L. H. VanBebber, S. Bhattacharya, P. Chandra, V. Keppens, D. Mandrus, M. A. McGuire,
B. C. Sales, A. S. Sefat, and J. Schmalian, Phys. Rev. Lett. {\bf 105}, 157003 (2010).

\bibitem{myi} M. Yi, D. H. Lu, J.-H. Chu, J. G. Analytis, A. P. Sorini, A. F. Kemper, B. Moritz, S.-K. Mo, R. G. Moore, M. Hashimoto, W. S. Lee, Z. Hussain, T. P. Devereaux, I. R. Fisher, Z.-X. Shen, Proc. Natl. Acad. Sci. U.S.A. {\bf 108}, 6878 (2011).

\bibitem{yzhang} Y. Zhang, C. He, Z. R. Ye, J. Jiang, F. Chen, 
M. Xu, Q. Q. Ge, B. P. Xie, J. Wei, M. Aeschlimann, 
X. Y. Cui, M. Shi, J. P. Hu, and D. L. Feng, Phys. Rev. B {\bf 85}, 085121 (2012).

\bibitem{myi12} M. Yi, D. H. Lu, R. G. Moore, K. Kihou, C.-H. Lee, A. Iyo, H. Eisaki, T. Yoshida, A. Fujimori, Z.-X. Shen, New J. Phys. {\bf 14}, 073019 (2012).


\bibitem{harriger} L. W. Harriger, H. Q. Luo, M. S. Liu, C. Frost, J. P. Hu, M. R. Norman, and P. C. Dai, Phys. Rev. B {\bf 84}, 054544 (2011).

\bibitem{kasahara} S. Kasahara,
 H. J. Shi,
 K. Hashimoto,
 S. Tonegawa,
 Y. Mizukami,
 T. Shibauchi,
 K. Sugimoto,
 T. Fukuda,
 T. Terashima,
 Andriy H. Nevidomskyy, and Y. Matsuda, Nature {\bf 486}, 382 (2012).

\bibitem{tmchuang}
T.-M. Chuang,
 M. P. Allan,
 Jinho Lee,
 Yang Xie,
 Ni Ni,
 S. L. Bud'ko,
 G. S. Boebinger,
 P. C. Canfield,
 J. C. Davis, Science {\bf 327}, 181 (2010).

\bibitem{allan} M. P. Allan, T.-M. Chuang, F. Massee, Yang Xie, Ni Ni, S. L. Bud'ko, G. S. Boebinger, Q. Wang, D. S. Dessau, P. C. Canfield, M. S. Golden, J. C. Davis, Nature Phys. {\bf 9}, 220 (2013).

\bibitem{pasupathy} E. P. Rosenthal, E. F. Andrade, C. J. Arguello, R. M. Fernandes, L. Y. Xing, X. C. Wang, C. Q. Jin, A. J. Millis, A. N. Pasupathy,
arXiv: 1307.3526.

\bibitem{fradkin} E. Fradkin, S. A. Kivelson, M. J. Lawler, J. P. Eisenstein, and A. P. Mackenzie, Annu. Rev. Condens. Matter Phys. {\bf 1}, 153 (2010).

\bibitem{jphu} J. P. Hu and C. K. Xu, Physica C {\bf 481}, 215 (2012).

\bibitem{fernandes1} R. M. Fernandes, A. V. Chubukov, J. Knolle, I. Eremin, and J. Schmalian, Phys. Rev. B {\bf 85}, 024534 (2012).

\bibitem{cclee} C. C. Lee, W. G. Yin, and W. Ku, Phys. Rev. Lett. {\bf 103}, 267001 (2009).

\bibitem{kruger} F. Kr$\rm \ddot{u}$ger, S. Kumar, J. Zaanen, and J. van den Brink, Phys. Rev. B {\bf 79}, 054504 (2009).

\bibitem{lv} W. C. Lv, J. S. Wu, and P. Phillips, Phys. Rev. B {\bf 80}, 224506 (2009).

\bibitem{ccchen} C.-C. Chen, J. Maciejko, A. P. Sorini, B. Moritz, R. R. P. Singh, and T. P. Devereaux, Phys. Rev. B {\bf 82}, 100504(R) (2010).

\bibitem{valenzeula} B. Valenzuela, E. Bascones, and M. J. Calder$\rm \acute{o}$n, Phys. Rev. Lett. {\bf 105}, 207202 (2010).

\bibitem{jhchu10} Jiun-Haw Chu, James G. Analytis, David Press, Kristiaan De Greve, Thaddeus D. Ladd, Yoshihisa Yamamoto, and Ian R. Fisher, Phys. Rev. B {\bf 81}, 214502 (2010).

\bibitem{ruff} J. P. C. Ruff, J.-H. Chu, H.-H. Kuo, R. K. Das, H. Nojiri, I. R. Fisher, and Z. Islam, Phys. Rev. Lett. {\bf 109}, 027004 (2012).

\bibitem{nakajima} M. Nakajima, S. Ishida, Y. Tomioka, K. Kihou, C. H. Lee, A. Iyo, T. Ito, T. Kakeshita, H. Eisaki, and S. Uchida, Phys. Rev. Lett. {\bf 109}, 217003 (2012).

\bibitem{ishida} S. Ishida, M. Nakajima, T. Liang, K. Kihou, C. H. Lee, A. Iyo, H. Eisaki, T. Kakeshita, Y. Tomioka, T. Ito, and S. Uchida, Phys. Rev. Lett. {\bf 110}, 207001 (2013).

\bibitem{olly} O. J. Lipscombe, L. W. Harriger, P. G. Freeman, M. Enderle, C. L. Zhang, M. Y. Wang, T. Egami, J. P. Hu, T. Xiang, M. R. Norman, and P. C. Dai, Phys. Rev. B {\bf 82}, 064515 (2010).

\bibitem{qureshi} N. Qureshi, P. Steffens, S. Wurmehl, S. Aswartham, B. B$\rm \ddot{u}$chner, and M. Braden, Phys. Rev. B {\bf 86}, 060410(R) (2012).

\bibitem{mengshu} Mengshu Liu, C. Lester, Jiri Kulda, Xinye Lu, Huiqian Luo, Meng Wang, S. M. Hayden, and Pengcheng Dai, Phys. Rev. B {\bf 85}, 214516 (2012).

\bibitem{clzhang} C. L. Zhang, M. S. Liu, Y. X. Su, L.-P. Regnault, M. Wang, G. T. Tan, Th. Br$\rm \ddot{u}$ckel, T. Egami, and P. C. Dai, Phys. Rev. B {\bf 87}, 081101(R) (2013).

\bibitem{hluo} Huiqian Luo, Meng Wang, Chenglin Zhang, Xingye Lu, Louis-Pierre Regnault, Rui Zhang, Shiliang Li, Jiangping Hu, and Pengcheng Dai, Phys. Rev. Lett. {\bf 111}, 107006 (2013).

\bibitem{collin} M. F. Collin, \textit{Magnetic Critical Scattering}, Chapter 8, Oxford series on neutron scattering 
in condensed matter, (Oxford University Press, USA, 1989).

\bibitem{jtpark12} J. T. Park, G. Friemel, T. Loew, V. Hinkov, Yuan Li, B. H. Min, D. L. Sun, A. Ivanov, A. Piovano, C. T. Lin, B. Keimer, Y. S. Kwon, and D. S. Inosov, Phys. Rev. B {\bf 86}, 024437 (2012).

\bibitem{matan} K. Matan, R. Morinaga, K. Iida, and T. J. Sato, Phys. Rev. B {\bf 79}, 054526 (2009).

\bibitem{clzhang13} Using time-of-flight neutron spectroscopy, we have collected spin wave data on NaFeAs throughout the entire Brillouin in the AF ordered state.  Based on Heisenberge Hamiltonian fits to the high-energy spin waves, we determine effective magnetic exchange couplings along different crystallographic directions.   

\bibitem{cwang} Chong Wang, Rui Zhang, Fa Wang, Huiqian Luo, L. P. Regnault, Pengcheng Dai, Yuan Li, arXiv: 1309.7553.

\bibitem{xylu13} X. Y. Lu, J. T. Park, R. Zhang, H. Q. Luo, A. H. Nevidomskyy, Q. Si, and P. C. Dai, 
(unpublished).

\bibitem{boothroyd}
P. Babkevich, B. Roessli, S. N. Gvasaliya, L.-P. Regnault, P. G. Freeman, E. Pomjakushina, K. Conder, and A. T. Boothroyd, Phys. Rev. B {\bf 83}, 180506(R) (2011).

\bibitem{prokes} K. Proke$\rm \check{s}$, A. Hiess, W. Bao, E. Wheeler, S. Landsgesell, and D. N. Argyriou, Phys. Rev. B {\bf 86}, 064503 (2012).

\bibitem{steffens} P. Steffens, C. H. Lee, N. Qureshi, K. Kihou, A. Iyo, H. Eisaki, and M. Braden, Phys. Rev. Lett. {\bf 110}, 137001 (2013).

\bibitem{leland09} L. W. Harriger, A. Schneidewind, S. L. Li, J. Zhao, Z. C. Li, W. Lu, X. L. Dong, F. Zhou, Z. X. Zhao, J. P. Hu, and P. C. Dai, Phys. Rev. Lett. {\bf 103}, 087005 (2009).

\end{thebibliography}
\end{document}